\documentclass[a4paper,11pt]{article}
\pdfoutput=1 % if your are submitting a pdflatex (i.e. if you have
\usepackage{jcappub} % for details on the use of the package, please

\usepackage[utf8]{inputenc}
\DeclareMathAlphabet\mathbfcal{OMS}{cmsy}{b}{n}

\title{Bayesian evidence for $\alpha$-attractor dark energy models}
\author[a,b,1]{Francisco X.~Linares Cede\~no, \note{Corresponding author.}}
\author[b]{Ariadna Montiel,} 
\author[b]{Juan Carlos Hidalgo,} 
\author[b]{Gabriel German}

\date{\today}
\affiliation[a]{Departamento de Física, DCI, Campus Le\'on, Universidad de Guanajuato,\\ 37150, León, Guanajuato, México.}
\affiliation[b]{Instituto de Ciencias F\'{\i}sicas, 
Universidad Nacional
Aut\'onoma de M\'exico, \\62210, Cuernavaca, Morelos, México}
\emailAdd{fran2012@fisica.ugto.mx}
\emailAdd{amontiel@icf.unam.mx}
\emailAdd{hidalgo@icf.unam.mx}
\emailAdd{gabriel@icf.unam.mx}

\usepackage{booktabs}
\usepackage{multirow}
\usepackage{natbib}
\usepackage{graphicx}
\usepackage{appendix}
\usepackage{yfonts}

\abstract{Dark energy models with tracker properties have gained attention due to the large range of initial conditions leading to the current value of the dark energy density parameter. A well-motivated family of these models are the so-called $\alpha$-\textit{attractors}, which show the late time behavior of a cosmological constant. In the present paper we perform a model-selection analysis of a variety of $\alpha$-attractor   potentials in comparison with a non-flat $\Lambda$CDM model. Specifically, we compute the Bayes Factor for the L-Model, the Oscillatory Tracker Model, the Recliner Model, and the Starobinsky Model,   while considering the non-flat $\Lambda$CDM as the base model. Each model is tested through a Bayesian analysis using observations relevant to the current accelerated expansion: we employ the latest SNe Ia data, combined with cosmic clocks, the latest BOSS release of BAO data, and the Planck Compressed 2018 data. The produced Markov Chains for each model are further compared through a Bayesian evidence analysis. From the latter we conclude that the Oscillatory Tracker Model is preferred by data (even if weakly) over the non-flat $\Lambda$CDM model. Our results also suggest at the L-model is the least favoured version of the $\alpha$-attractor models considered.}

\begin{document}

\maketitle

\section{Introduction}
The accelerated expansion of the Universe constitutes a paradigm break in our understanding of its large scale dynamics. The current cosmic acceleration was first discovered through observations of nearby and distant type Ia Supernova at the end of 90s \cite{Riess:1998cb,Perlmutter:1998np}. The data from further observations, e.g. the Cosmic Microwave Background anisotropies (CMB) \cite{Aghanim:2018eyx} or the large scale structure \cite{Eisenstein:2005su,Blake2012MNRAS,WiggleZPhysRevD86,Kazin:2014qga,Beutler:2015tla}, shows consistency with this hypothesis. The current accelerated expansion is attributed to the so-called \textit{Dark Energy} (DE), constituting $\sim 70\%$ of the total energy density of the Universe. So far, the $\Lambda$CDM model consisting of a cosmological constant $\Lambda$ plus cold dark matter is the most accepted cosmological model. However, theoretical inconsistencies (e.g. fine tuning and the cosmic coincidence
problems \cite{Weinberg1989}) motivate cosmologist to reach for alternative (dynamical) origin of DE. Among others, models such as \textit{Quintessence} (\cite{Linder:2007wa,Tsujikawa:2013fta,Chiba:2012cb,Durrive:2018quo}), \textit{Phantom Dark Energy} (\cite{Caldwell:1999ew,Caldwell:2003vq,Nojiri:2005sx,Ludwick:2018omw}), and \textit{Quintom Dark Energy} (\cite{Feng:2006ya,Setare:2008sf,Leon:2018lnd}) consider a scalar field as the responsible of the DE dynamics, while the law of gravitation is given by \textit{General Relativity} (GR). An alternative approach is the  extension of the gravity theory beyond GR, in a strategy generically dubbed \textit{Modified Gravity}~\cite{Lobo:2008sg,Tsujikawa:2010sc,Li:2011sd,Clifton:2011jh,Dimitrijevic:2012kb,Brax:2015cla,Joyce:2016vqv,Jaime:2018ftn}.

In the context of Dark Energy, cosmological models with tracker properties have gained attention since the scalar field reaches the present value of the DE density from a wide range of initial conditions~\cite{Zlatev:1998tr} thereby  alleviating the coincidence problem of an extremely small cosmological constant. The present work looks at a family of tracking Dark Energy models known as $\alpha$-\textit{attractors}, modeled through a scalar field $\phi$ with a Quintessence-like behaviour. 

The $\alpha$-\textit{attractors} were originally studied in a superconformal approach to the description of inflationary models \cite{Kallosh:2013wya}, \cite{Kallosh:2013xya}. This approach is based on previous studies of cosmology within $N=1$ gauge theory superconformally coupled to supergravity \cite{Kallosh:2000ve}. It is observed \cite{Kallosh:2013hoa} that a large class of inflationary models with very similar predictions share an attractor point
\begin{equation}
    1-n_s=2/N, \quad r=12\alpha /N^2\, ,
      \label{point} 
\end{equation}
where $N$ is the number of e-folds of inflation. Expressions \eqref{point} are given in the leading approximation in $1/N$. The $\alpha$-attractor models were further studied as inflationary models by \cite{Kallosh:2013yoa,Kallosh:2016sej,Odintsov:2016vzz,Ueno:2016dim,Kumar:2015mfa,Eshaghi:2016kne}. It was shown that the scalar field can be described in a canonical form with a potential given by
\begin{equation}
    V(\phi)\propto F\left[\tanh \left(\frac{\kappa \phi}{\sqrt[]{6\alpha}}\right)\right]\, ,
\end{equation}
i.e., with the potential $V(\phi)$ as a function $F$ of the hyperbolic tangent of the scalar field. In the above expression, $\kappa = \sqrt{8\pi G}$, and $\alpha$ is a free parameter inversely proportional to the curvature of the inflaton K\"ahler manifold~\cite{Kallosh:2013yoa}. Due to the tracker properties of these models, $\alpha$-attractor potentials have been employed to study the late time Universe, resulting in an optimal description of the current accelerated expansion  \cite{Linde:2015uga,Linder:2015qxa,Kallosh:2015lwa,Carrasco:2015pla,Scalisi:2015qga,Shahalam:2016juu,Bag:2017vjp,Dimopoulos:2017zvq,Dimopoulos:2017tud,Garcia-Garcia:2018hlc}. Furthermore, in \cite{Akrami:2017cir} quintessential $\alpha$-attractor inflationary models are extensively studied. In particular, $\alpha$-attractor models where a single field plays the double role of the inflaton and the quintessence and models where inflaton and quintessence are described by two different fields.

The realisations of the $\alpha$-attractor family studied in \cite{Bag:2017vjp} are,
\begin{eqnarray}
\left(\frac{V(\phi)}{\alpha c^2}\right) = \left\{
        \begin{array}{ll}
            \left[\tanh\left(\frac{\kappa\phi}{\sqrt{6\alpha}}\right)\right]^{-1} = \coth\left(\frac{\kappa\phi}{\sqrt{6\alpha}}\right) \, , & \quad {\rm{L - Model}}\, , \\
            \left[1 - \tanh^2\left(\frac{\kappa\phi}{\sqrt{6\alpha}}\right)\right]^{-1/2}= \cosh\left(\frac{\kappa\phi}{\sqrt{6\alpha}}\right) \, , & \quad {\rm{Oscillatory\ Tracker\ Model}}\, , \\
            \left[1 + \tanh\left(\frac{\kappa\phi}{2\sqrt{6\alpha}}\right)\right]^{-1} = \left( 1 + e^{-\frac{\kappa\phi}{\sqrt{6\alpha}}} \right)\, , & \quad {\rm{Recliner\ Model}} \, , \\ \tanh^2\left(\frac{\kappa\phi}{\sqrt{6\alpha_1}}\right)\cosh\left(\frac{\kappa\phi}{\sqrt{6\alpha_2}}\right)\, , \quad \ \alpha_1 \ll \alpha_2\, , & \quad {\rm{Margarita\ potential}} \, ,
        \end{array}
    \right.
    \label{potentials}
\end{eqnarray}
where $\alpha$ and $c$ are the characteristic parameters of the $\alpha$-attractor models. On the other hand, Ref.~\cite{Garcia-Garcia:2018hlc} proposes a generalization to the $\alpha$-attractor potentials written as
\begin{equation}
    V(\chi) = \alpha c^2 \frac{\chi^p}{\left( 1 + \chi \right)^{2n}}\, ,\quad {\rm with}\quad \chi \equiv \tanh\left( \kappa \phi/\sqrt{6\alpha} \right)\, ,
    \label{potential_chi}
\end{equation}
with $p$ and $n$ indices which values specify a particular potential. For instance, for the combination $(p=-1,n=0)$ the L-Model is recovered, while for $(p=0,n=1/2)$ we obtain the Recliner model. Moreover, it is straightforward to derive a \textit{Starobinsky-like} potential from Eq.~\eqref{potential_chi}; in fact, the Starobinsky model corresponds to $\alpha=1$ and $(p=2,n=1)$.

The success of $\alpha$-attractor potentials \eqref{potentials} and~\eqref{potential_chi} in describing the late Universe poses the question of whether this quintessence family performs better than the $\Lambda$CDM model. Comparisons so far have analyzed the cosmological evolution of these models, their asymptotic behavior at early and late times, the Matter and Temperature Power Spectra, and even the Bayesian likelihood of the different parameters of the model (see in particular Refs.~\cite{Bag:2017vjp,Garcia-Garcia:2018hlc}).

It is the main goal of this paper to  compare the performance of $\alpha$-attractor models with that of the $\Lambda$CDM model via the statistical information provided by the observations relevant to the late-time expansion. In order to facilitate the comparison, we place all models under the same parametrization by generalizing the potential \eqref{potential_chi} as follows
\begin{equation}
    V(\chi) = \alpha c^2 \frac{\chi^p}{\left( 1 + A\chi^q \right)^{2n}}\, .
    \label{gen_potential}
\end{equation}

\noindent The values of $p\, , n\, , A\, ,$ and $q$ that reproduce each of the potentials considered in this work are shown in Table~\ref{pqnA}. In consequence, the only free parameters, common to all the $\alpha$-attractor models of interest, are the constants $\alpha$ and $c\, .$
\begin{table}[h!]
\centering
\begin{tabular}{ccccc}
\hline
\hline
Potential & $p$    & $n$ & $A$ & $q$  \\
\hline
\hline
L-Model& $-1$ & $0$  & $1$ & $1$  \\

Oscillatory Tracker Model& $0$ & $1/4$  & $-1$ & $2$ \\

Recliner Model& $0$ & $1/2$  & $1$ & $1$  \\

Starobinsky Model& $2$ & $1$  & $1$ & $1$  \\
 \hline
\end{tabular}
\caption{Values of $p\, , n\, , A\, ,$ and $q$ to set the particular $\alpha$-attractor potential.}
\label{pqnA}
\end{table}

In this work we formally compare $\alpha$-attractor models against the $o\Lambda$CDM model (which refers to $\Lambda$CDM with an allowed contribution of curvature $\Omega_k$), by computing the \textit{Bayes Factor} $B$, which is widely acknowledged as the most reliable statistical tool for model comparison. The outline of the present paper is the following. In Section~\ref{BI}, we introduce the basics of Bayesian Inference making emphasis on the Bayesian Evidence, which is the crucial tool to calculate in model comparison. In Section~\ref{SMOS} we describe the observational data  used to constrain each of the $\alpha$-attractor potentials. We consider three sets of local observations (Type Ia Supernovae, Hubble data, and Baryon Acoustic Oscillations) and one cosmological dataset (CMB). Our results are presented in Section~\ref{results}, where we show the cosmological evolution of the DE density parameter for the $o\Lambda$CDM model and for the $\alpha$-attractor models shown in Table~\ref{pqnA}. The same section presents the posterior probabilities for the particular case of the Oscillatory Tracker Model, and the main result of this work: the Evidence and the Bayes factor for all models. In Section~\ref{disc} we discuss our results and draw conclusions from them.

\section{Bayesian inference}\label{BI}

According to Bayes' theorem, the probability of a model $M$ with a set of parameters $\Theta$, in light of the observed data $D$, is given by the  \textit{Posterior} $\mathcal{P}$: 
\begin{equation}
    \mathcal{P}(\Theta \mid D,M)= \frac{\mathcal{L}(D\mid \Theta,M)\Pi(\Theta\mid M)}{\mathcal{E}(D\mid M)},
    \label{eq:bayes}
\end{equation}

\noindent where $\mathcal{L}$ is the Likelihood function, $\Pi$ represents the set of  Priors, containing the \textit{a priori} information about the parameters of the model. $\mathcal{E}$ is the so-called Evidence, to which we pay particular attention.

For a given model $M$, the Bayesian evidence $\mathcal{E}$ (hereafter simply the evidence) is the normalizing constant in the right hand side of Eq. (\ref{eq:bayes}). It normalises the area under the posterior $\mathcal{P}$ to unity, and is given by
\begin{equation}
  \mathcal{E}(D\mid M)= \int d\Theta \mathcal{L}(D|\Theta,M)\Pi(\Theta|M)\, . 
  \label{eq:evidence}
\end{equation}

The evidence can be neglected in model fitting, but it becomes important in model comparison\footnote{There are several approaches to perform a model selection statistics. For instance, there are the \textit{Akaike Information Criterion} (AIC)~\cite{akaike1974new} and the \textit{Bayesian Information Criterion} (BIC)~\cite{Schwarz:1978tpv}, in which the maximum likelihood $\mathcal{L}_{max}$ is used to minimize the information of the model. The AIC includes the number of parameters of the model, whereas the BIC adds the number of data points as well. However, all these approaches are not Bayesian since they ignore the Prior, one of the main ingredients of the Bayes' theorem~\eqref{eq:bayes}.}. In fact, when comparing two different models $M_1$ and $M_2$ using Bayes' theorem \eqref{eq:bayes}, the ratio of posterior probabilities of the two models $\mathcal{P}_1$ and $\mathcal{P}_2$ will be proportional to the ratio of their evidences, this is
\begin{equation}
    \frac{\mathcal{P}_1(\Theta_1 \mid D,M_1)}{\mathcal{P}_2(\Theta_2 \mid D,M_2)} = \frac{\Pi_1(\Theta_1|M_1)}{\Pi_2(\Theta_2|M_2)}\frac{\mathcal{E}_1(D\mid M_1)}{\mathcal{E}_2(D\mid M_2)}\, .
\end{equation}

This ratio between posteriors leads to the definition of the \textit{Bayes Factor} $B_{12}$, which in logarithmic scale is written as
\begin{equation}
    \log B_{12} \equiv \log \left[\frac{\mathcal{E}_1(D\mid M_1)}{\mathcal{E}_2(D\mid M_2)}\right] = \log \left[\mathcal{E}_1(D\mid M_1)\right] - \log \left[\mathcal{E}_2(D\mid M_2)\right] \, .
    \label{BF}
\end{equation}

If $\log B_{12}$ is larger (smaller) than unity, the data favours model $M_1$ ($M_2$). To assess the strength of the evidence contained in the data, Jeffreys \cite{Jeffreys1961} introduces an empirical scale, see Table~\ref{Jeffreyscale}. For a comprehensive review of Bayesian model selection, we refer the reader to \cite{Trotta:2008qt}. 
\begin{table}[h!]
\centering
\begin{tabular}{cc}
\hline
\hline
$2\ln{B_{12}}$    & Strength  \\
\hline
\hline
$<$ 0      & Negative (support $M_2$)      \\
0 - 2.2 & Weak       \\
2.2 - 6 & Positive     \\
6 - 10      & Strong     \\
$>10$ & Very strong \\
\hline
\end{tabular}
\caption{Jeffrey's scale to quantify the strength of evidence for a corresponding range of the Bayes factor. We follow the convention of \cite{Kass:1995loi,10.2307/2337598} in presenting a factor of two with the natural logarithm of the Bayes factor.}
\label{Jeffreyscale}
\end{table}

Here we calculate the evidence for each realization of the $\alpha$-attractor model $\mathcal{E}_{\alpha}$, as well as the evidence for the $o\Lambda$CDM model $\mathcal{E}_{\Lambda}$, to then compute the Bayes factor $B_{\alpha \Lambda}$ according to Eq.~\eqref{BF}. This will allow us to assess the viability of the $\alpha$-attractor models in describing the current accelerated expansion of the Universe, in comparison to the cosmological constant $\Lambda$.

 A variety of computational techniques have been developed to derive the evidence~\eqref{eq:evidence}: \textit{Laplace Approximation} (LA)~\cite{de1958asymptotic,tierney1986accurate,bleistein1986asymptotic}, \textit{Variational Bayes} (VB)~\cite{mackay1997ensemble,vsmidl2006variational}, \textit{Nested Sampling} (NS)~\cite{Skilling:2006gxv}, \textit{Importance Sampling} (IS)~\cite{2008arXiv0801.3887C,2009AIPC.1193..251R}. Particularly for NS and IS sophisticated numerical codes  compute the evidence for cosmological models (\textsc{multinest}~\cite{Feroz:2007kg,Feroz:2008xx,Feroz:2013hea}, \textsc{polychord}~\cite{Handley:2015fda,2015MNRAS.453.4384H}, and as two of us have previously tested in \cite{Barbosa-Cendejas:2017pbo}). Such codes perform the integration of~\eqref{eq:evidence} over all the parameter space, with the capacity of  handling such multidimensional integral for multi-peaked likelihoods. Recently, another proposal to calculate the evidence has been reported in Ref.~\cite{Heavens:2017afc}, where the authors use the fact that the unnormalized posterior $\tilde{\mathcal{P}}(\Theta\mid D,M)$ is proportional to the number density $n(\Theta\mid D,M)$, that is, $\tilde{\mathcal{P}}(\Theta\mid D,M) = a\ n(\Theta\mid D,M)\,$. Since the number density is given by
\begin{equation}
    n(\Theta\mid D,M) = N\ \mathcal{P}(\Theta\mid D,M) = N\ \frac{\tilde{\mathcal{P}}(\Theta\mid D,M)}{\mathcal{E}(D \mid M)},
\end{equation}

\noindent where $N$ is the lenght of the chain, then,
\begin{equation}
    \quad \mathcal{E}(D \mid M) = a\ N\, .
\end{equation}
 
\noindent Therefore, once the proportionality constant $a$ is determined, it is possible to calculate the evidence $\mathcal{E}$ directly from the MCMC chains. The software developed in~\cite{Heavens:2017afc}, called \textsc{MCEvidence}, is designed to compute the Bayesian evidence from MCMC sampled posterior distributions\footnote{This code has been successfully employed to calculate the evidence for extensions of the $\Lambda$CDM model~\cite{Heavens:2017hkr}, extra dimensional extension of GR~\cite{Kouwn:2017qet}, the effect on cosmology due to different models of neutrinos~\cite{Long:2017dru}, deviation of $\Lambda$CDM model considering oscillating DE parametrization~\cite{Pan:2017zoh}, and to test different models of reionizaion scenarios for future 21cm observations~\cite{Binnie:2019dwt}, among other applications.}. The code takes the $k-$th nearest-neighbor distances in parameter space with distances computed using the Mahalanobis distance, where the inverse covariance matrix estimated from the MCMC chains defines the metric, to estimate the Bayesian evidence from the MCMC samples provided by the chains. We shall quote results for $k=1$, which seems to be the most accurate choice \cite{Heavens:2017afc}. To reduce parameter correlations, one may wish to thin the chains. However, the effect on the Bayes factor is small as was shown in \cite{Heavens:2017afc}.

In this paper, we use the \textsc{MCEvidence} code to compute the evidences from the collection of our chains for the $\alpha$-attractor models and the $o\Lambda$CDM model. In the next section, we briefly discuss the set of data we used to test our models.

\section{Computational tools and Observational samples}\label{SMOS}

Before addressing the observational data we have employed, it is worth mentioning that we have used the Boltzmann code \textsc{class} \cite{Lesgourgues:2011re} to run the background evolution of the $\alpha$-attractor models. This allowed us to explore the set of initial conditions for each potential, and we found that as expected a large amount of values of $\alpha\, ,c\, ,$ and $\phi_i$ are able to reproduce the DE behavior. We used the constant $\alpha$ as a shooting parameter to find the proper initial conditions for the scalar field evolution; for a shooting range between $\alpha^{sh} = 10^{-8} - 10^{-6}$, the values for the parameters are $\alpha = 0 - 10\, , c = 10^{-7} - 10^{-5}\, ,\phi_i = 0 - 10\, ,$ and $\dot{\phi}_i = 0\, .$ Hereafter we report values of the shooting parameter $\alpha^{sh}$ only. Note that the initial condition for the scalar field velocity is set to zero for all the values of $(\alpha\, ,c\, , \phi_i)$, since at early times the Hubble friction will freeze the field, as it starts for a nearly flat plateau \cite{Garcia-Garcia:2018hlc}. The Starobinsky potential is recovered for $\alpha=1$; however, for this particular case we allow $\alpha$ to take values between $0$ and $1$.

On the other hand, to find the high confidence region of the parameter space of the $\alpha$-attractor models given a set of observational data, we use the Markov Chain Monte Carlo (MCMC) method. In particular, we used the publicly available \textsc{monte python} package \cite{Audren:2012wb}, which is a cosmological parameter estimator linked to \textsc{class}. The code employs the Metropolis-Hastings algorithm \cite{Metropolis:1953am,Hastings:1970aa} for sampling, and it computes the Bayesian parameter inference of the posteriors with the convergence test given by the Gelman-Rubin criterion $R$ \cite{gelman1992}, where we require $R-1<10^{-3}$ for all our chains. The chains generated with the MCMC method will be also useful to calculate the bayesian evidence, as we will show in Section~\ref{results}. 

By using the publicly available \textsc{monte python} \cite{Audren:2012wb} package, we perform a likelihood analysis in which we minimize the $\chi^2$ function thus obtaining the best fit of model parameters from observational data (see Section~\ref{results} for more details). This minimization is equivalent to maximizing the likelihood function $\mathcal{L}(\theta) \propto \exp [-\chi^2(\theta)/2]$
where $\theta$ is the vector of model parameters and the expression for $\chi^2(\theta)$ depends on the
used dataset. In what follows we briefly describe the observational samples.

\subsection{Type Ia Supernovae (SNe Ia)}

We used the SNe Ia data from the Pantheon compilation \cite{Scolnic:2017caz}. This set is made of 1048 SNe covering the redshift range $0.01<z< 2.26$. As is usual, the likelihood from SNe Ia data is constrained from the standard $\chi^2$ statistics given by
\begin{equation}
\chi^2_{SN}= \mathrm{\Delta} \mathrm{\mu} \cdot \mathrm{C^{-1}} \cdot \mathrm{\Delta} \mathrm{\mu},
\label{Eq:CSNIa}
\end{equation}
where $\mathrm{C}$ is the full systematic covariance matrix and $\mathrm{\Delta} \mathrm{\mu}=\mathrm{\mu_{theo}}
-\mathrm{\mu_{obs}}$ is the vector of the differences between the observed and theoretical value of the observable quantity for SNe Ia, the distance modulus, $\mathrm{\mu}$, defined as
\begin{equation}
\mu(z,\theta)= 5 \log_{10} \left[d_L(z,\theta) \right] + \mu_0,
\label{Eq:muSN}
\end{equation}
where $d_L(z,\theta) $ is the dimensionless luminosity distance given by
\begin{equation}
d_L(z,\theta)= (1+z) \int_0^z \frac{dz'}{E(z',\theta)},
\end{equation}
with $E(z,\theta)=H(z,\theta)/H_0$ the dimensionless Hubble function, $H_0$ the Hubble constant
and $\theta$ the free parameters of the cosmological model. In Eq. (\ref{Eq:muSN}), $\mu_0$ depends on both the absolute magnitude and the Hubble constant which are correlated. In our analysis, the absolute magnitude is taken as the  nuisance parameter.

\subsection{Baryon Acoustic Oscillations (BAO)}

We used data from the SDSS-III Baryon Oscillation Spectroscopic Survey (BOSS) DR12 \cite{Alam:2016hwk}, which provides measurements of both Hubble parameter $H(z_i)$ and the comoving angular diameter distance $d_A(z_i)$, at three separate redshifts, $z_i=\{ 0.38, 0.51, 0.61\}$ expressed as 
\begin{equation}
d_{A}(z) \frac{r^{fid}_{s}(z_{d})}{r_{s}(z_{d})}, \qquad \mathrm{and} \qquad H(z) \frac{r_{s}(z_{d})}{r^{fid}_{s}(z_{d})} \, ,
\end{equation}
where $r_{s}(z_{d})$ is the sound horizon evaluated at the dragging redshift $z_{d}$; and $r^{fid}_{s}(z_{d})$ is the same sound horizon but calculated for a given fiducial cosmological model used, being equal to $147.78$ Mpc \cite{Alam:2016hwk}.

The other two BAO data that we use, 6dF Galaxy Survey \cite{2011MNRAS.416.3017B} and SDSS Data Release 7 Main Galaxy Sample \cite{Ross:2014qpa}, are lower signal-to-noise and can only tightly constrain the spherically averaged combination of transverse and radial BAO modes, $$D_V(z) \equiv \left[cz(1+z)^2D_A^2(z) /H(z)\right]^{1/3}$$ These constraints are at respective redshifts $z = 0.106$ (6dF) and $z=0.15$ (SDSS MGS).\\

Thus, the corresponding $\chi_{BAO}^2$ for Baryon Acoustic Oscillations (BAO) is given by
\begin{equation}
\chi^2_{BAO}= \mathrm{\Delta} \mathbfcal{F}^{BAO} \cdot \mathbf{C}_{BAO}^{-1} \cdot \mathrm{\Delta} \mathbfcal{F}^{BAO}    ,
\label{Eq:chiBAO}
\end{equation}
where $\mathrm{\Delta} \mathbfcal{F}^{BAO}=\mathcal{F}_{theo}-\mathcal{F}_{obs}$ is the difference between the observed and theoretical value of the observable quantity for BAO which can be different depending on the considered survey and $\mathbf{C}^{-1}_{BAO}$ is the respective inverse covariance matrix.

\subsection{Observational Hubble Data (OHD)}

We have used the cosmic chronometers approach which was initially proposed by \cite{Jimenez_2002}. It provides an independent technique to constrain the expansion history of the Universe $H(z)$ from the differential evolution of massive and passive early-type galaxies \cite{Jimenez_2002,10.1093/mnrasl/slv037}. So far the main complication of the cosmic chronometers approach, { also referred as observational Hubble parameter data (OHD) \cite{Yi:2007gu}}, is the number of data points available in comparison
with SNe Ia luminosity distance data. However, many authors have demonstrated OHD can be competitive with SNe Ia and BAO datasets in constraining cosmological parameters since it imposes direct constraints on the expansion rate of the Universe at different epochs \cite{2011ApJ...730...74M,Moresco:2012by,2012PhRvL.109q1301Z,RiemerSorensen:2012ve}.

We used 30 data points in the redshift range $0.07 \leq z \leq 1.965$ reported in \cite{Moresco:2016mzx}. In this case, the $\chi^2_{OHD}$ estimator is defined as
\begin{equation}
\chi^2_{OHD}= \sum^{30}_{j=1} \frac{\left[ H_{th}(z_j,\mathbf{\theta})
-H_{obs}(z_j)\right]^2}{\sigma^2_{H_{obs}}(z_j)},
\label{Eq:COHD}
\end{equation}
with $\sigma^2_{H_{obs}}$ the measurement variances and $\mathbf{\theta}$ the vector of the free parameters of the cosmological model.

\subsection{Cosmic Microwave Background (CMB)}

Instead of the full data of the CMB anisotropies, we used CMB data in the condensed form of shift parameters (also known as distance priors) reported in \cite{Chen:2018dbv}, which were derived from the last release of the Planck results \cite{Aghanim:2018eyx}. Clearly, the analysis proceeds much faster in this way than by performing an analysis involving the full CMB likelihood. It is worth mentioning that this compressed likelihood of CMB can be used to study models with either non-zero curvature or a smooth DE component, as in our case, but not for modifications of gravity \cite{Mukherjee:2008kd,Ade:2015rim}. Indeed, the $\alpha$-attractor models lie among the smooth dark energy models, meaning they are phenomenologically too close to $\Lambda$CDM, as it was also noted in \cite{Garcia-Garcia:2018hlc}. So, we take advantage of the shift parameters, $(R,l_A,\Omega_bh^2,n_s)$ which provide an efficient summary of CMB data as far as dark energy constraints are concerned (as it has been argued in several works \cite{Kosowsky:2002zt,Wang:2007mza,Mukherjee:2008kd,Ade:2015rim}).

The first two quantities in the vector $(R,l_A,\Omega_bh^2,n_s)$ are defined as 
\begin{equation}
R \equiv \sqrt{\Omega_m H_0^2} \frac{d_A(z_{*})}{c},
\end{equation}
\begin{equation}
l_A\equiv \pi \frac{d_A(z_{*})}{r_s(z_{*})},
\end{equation}
where $d_A(z)$ is the comoving angular diameter distance and $r_s(z)$ is the comoving size of the sound horizon, both evaluated at photon-decoupling epoch $z_{*}$. 

The corresponding $\chi^2$ for the CMB is

\begin{equation}
\chi^2_{CMB}= \mathrm{\Delta} \mathbfcal{F}^{CMB} \cdot \mathbf{C}_{CMB}^{-1} \cdot \mathrm{\Delta} \mathbfcal{F}^{CMB}    ,
\label{Eq:chiCMB}
\end{equation}
where $\mathcal{F}^{CMB}=(R,l_A,\Omega_bh^2,n_s)$ is the vector of the shift parameters and $\mathbf{C}^{-1}_{CMB}$ is the respective inverse covariance matrix. The mean values for these shift parameters as well as their standard deviations and normalized covariance matrix are taken from Table 1 of \cite{Chen:2018dbv}.

\section{Results}\label{results}

\begin{figure}[h!]
\centering
\includegraphics[scale=0.95]{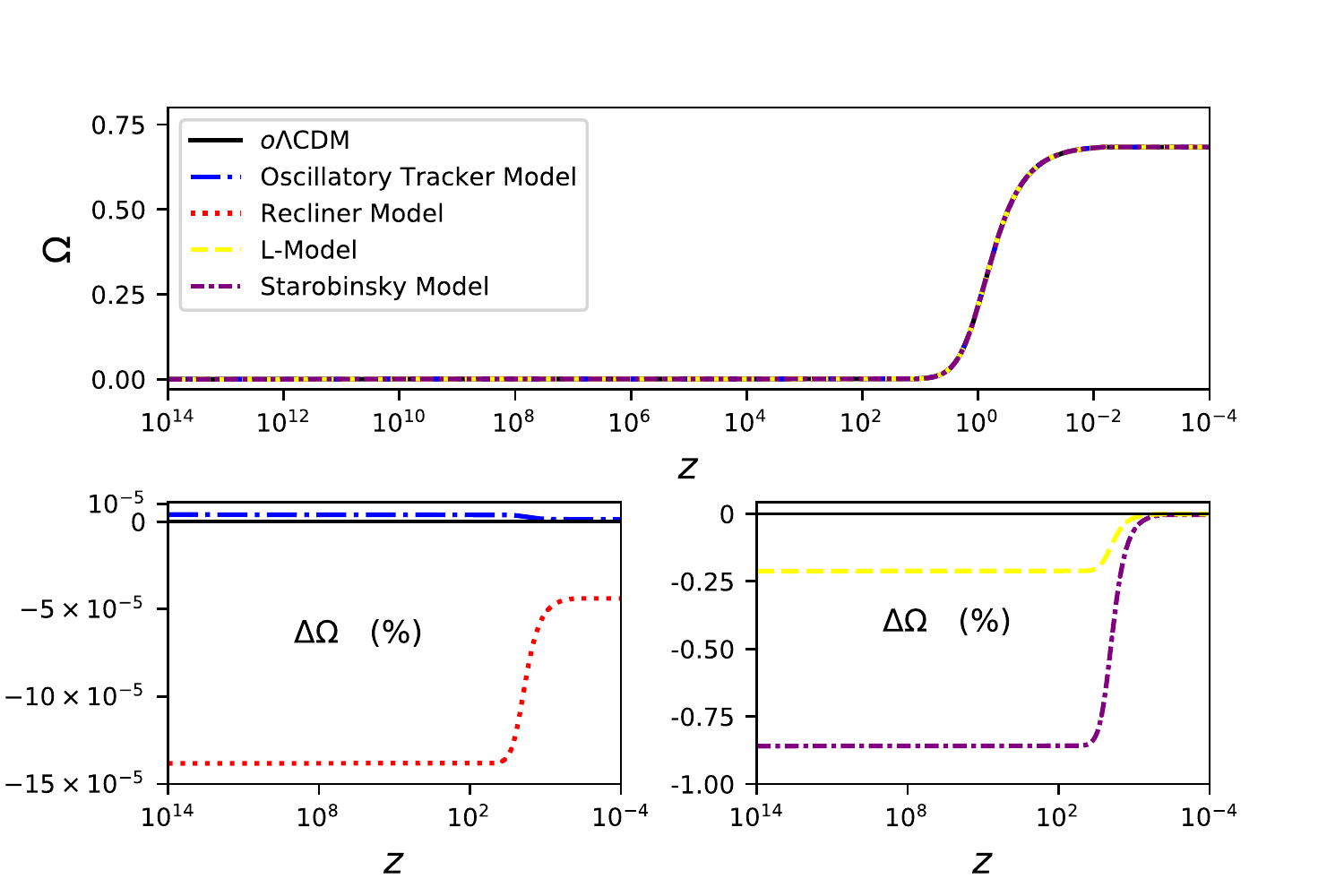}
\caption{Top panel: cosmological evolution of the DE density parameter $\Omega$ for $o\Lambda$CDM, and for all the $\alpha$-attractor potentials under study. Bottom panel: relative differences between $\Omega_{\Lambda}$ and each of the $\Omega_{\alpha}$ energy densities. See the text for more details.}
\label{omegas}
\end{figure}

As was mentioned above, the behavior of the $\alpha$-attractor models in the late Universe has been studied elsewhere \cite{Bag:2017vjp,Garcia-Garcia:2018hlc}, and it is not our goal to analyze the cosmological evolution for all potentials in \eqref{gen_potential}. Nonetheless, we want to emphasize that such potentials offer the same qualitative features as those of a cosmological constant. As presented in the first part of  Section~\ref{SMOS}, we have evolved the models using the \textsc{class} code. As can be seen in Figure~\ref{omegas} (top panel), the evolution of the DE density parameter for $o\Lambda$CDM ($\Omega_{\Lambda}$), as well as for our $\alpha$-attractor potentials ($\Omega_{\alpha}$), is strikingly similar. However, when looking at the relative differences between models (bottom panel), defined as $\Delta \Omega \equiv \left( \frac{\Omega_{\Lambda} - \Omega_{\alpha}}{\Omega_{\Lambda}} \right)*100\, $, we observe that tiny variations are present at sub-percentile level.

Table~\ref{paramfile} shows the input for the mean values, priors, and standard deviation for all the cosmological parameters $\Theta$ under consideration, which are given by:
\begin{equation}
    o\Lambda {\rm{CDM}}: \Theta_{\Lambda} = \left[ \Omega_{cdm}\, , \Omega_k \right]\, ,\quad \alpha-{\rm{attractors}}: \Theta_{\alpha} = \left[ \Omega_{cdm}\, , \Omega_k\, , \log \alpha\, , \log c \right]\,.
    \label{cosmoparam}
\end{equation}
Since we have obtained basically the same results for all the scalar field initial conditions $(\phi_i\, ,\dot{\phi}_i) = (0 \,\textrm{to}\, 10 , 0)$, we choose the initial conditions as $(\phi_i\, ,\dot{\phi}_i) = (10 , 0)\, .$ In the particular case of the Starobinsky potential, we have allowed $\alpha$ to take values between $0$ and $1$. Such values were achieved through the shooting method for $\log c= (0.27\, , 0.30)$, and $\log \alpha^{sh}= (-7\, , -5)\, .$ The set of priors for the MCMC analysis is shown in Table \ref{paramfile}.
\begin{table}[h!]
\centering
\begin{tabular}{cccccc}
\hline
\hline
& parameter    & mean & min prior & max prior & Std. Dev. \\
\hline
\hline
$o\Lambda$CDM & $\Omega_{cdm}$ & 0.2562  & 0.1 & 0.5 & 0.008 \\
& $\Omega_k$ & 0  & -0.2 & 0.2 & 0.005 \\
\hline
$\alpha$-attractors & $\Omega_{cdm}$ & 0.2562  & 0.1 & 0.5 & 0.008 \\
 & $\Omega_k$ & 0  & -0.2 & 0.2 & 0.005 \\
 & $\log \alpha^{sh}$ & -7 & -8 & -6 & 0.01 \\
 & $\log c$ & -2 & -4 & -0.38 & 0.05 \\
\hline
\end{tabular}
\caption{Input for $o\Lambda$CDM and $\alpha$-attractor parameters to generate the MCMC. We have setted the initial value of the scalar field and its velocity as $\phi_i = 10\, ,$ and $\dot{\phi}_i = 0$ respectively. For the Starobinsky potential, the mean, minimum and maximum priors for (the shooting parameter) $\alpha^{sh}$ and $c$ were set (in logarithmic scale) respectively to $\log \alpha^{sh}= (-6\, ,-7\, , -5)\, ,$ $\log c= (0.28\, ,0.27\, , 0.30)$.}
\label{paramfile}
\end{table}

\begin{figure}
\centering
\begin{minipage}{.5\textwidth}
  \centering
  \includegraphics[width=1.0\linewidth]{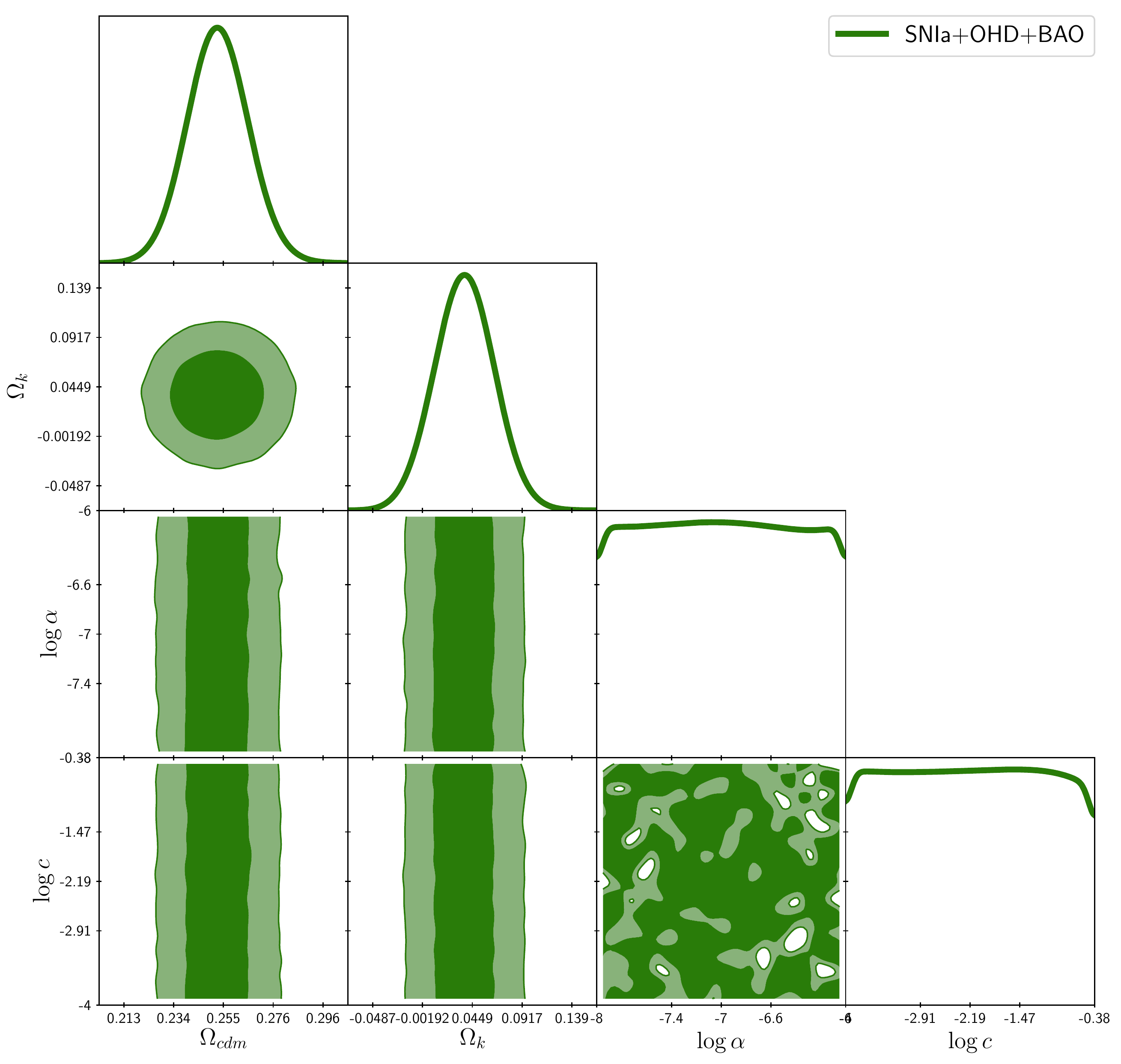}
  \label{fig:test1}
\end{minipage}%
\begin{minipage}{.5\textwidth}
  \centering
  \includegraphics[width=1.0\linewidth]{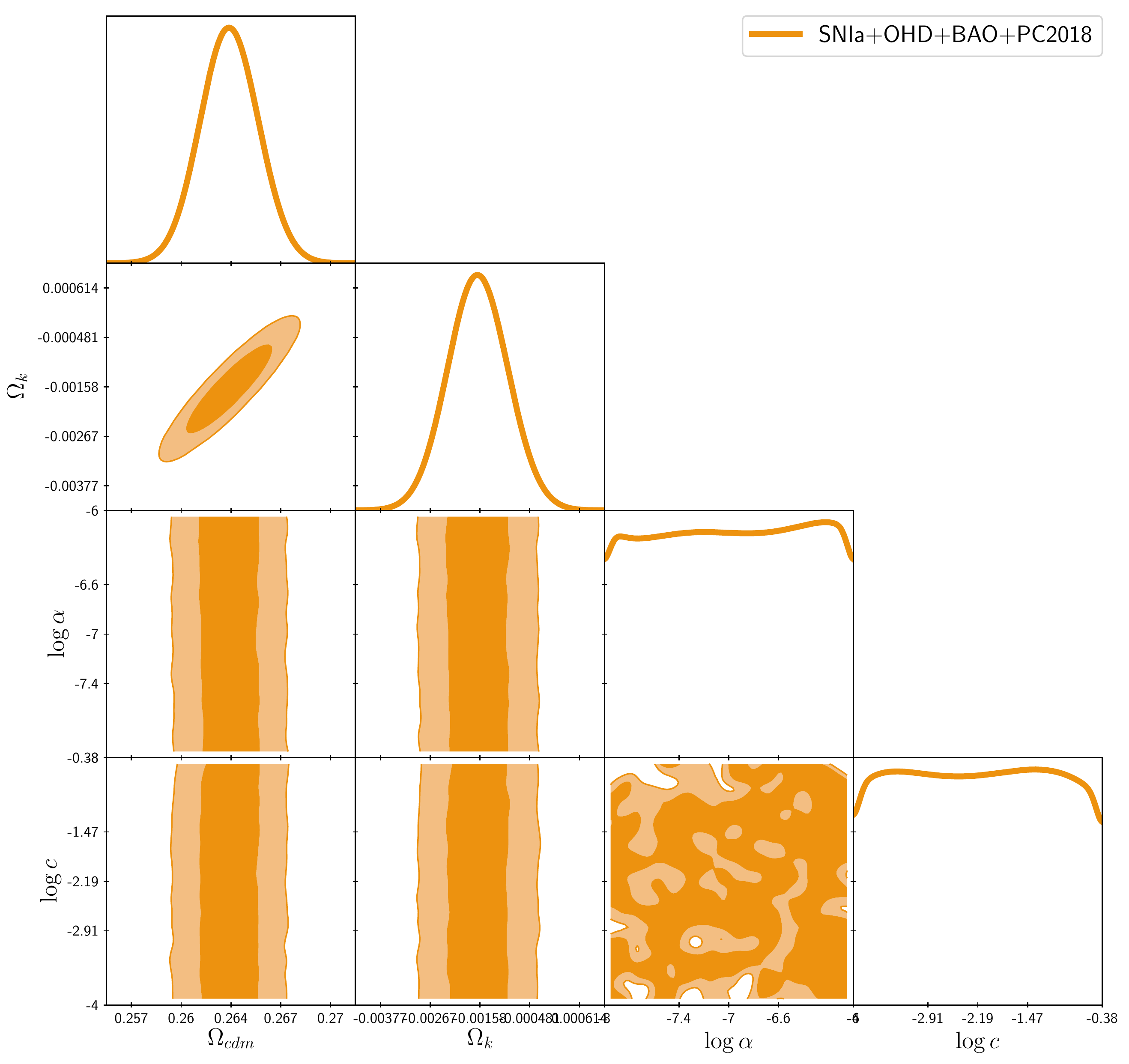}
  \label{fig:test2}
\end{minipage}
\caption{Posterior distributions for the Oscillatory Tracker Model. The parameters of the $\alpha$-attractor potential are presented in logarithmic scale $\log \alpha\, , \log c$ (bare in mind that the range of $\alpha$ plotted here corresponds to the shooting values). Left: parameter estimation using local observations data (SNe Ia+OHD+BAO). Right: parameter estimation adding data from Planck Compressed 2018 (PC2018).}
\label{posterior}
\end{figure}

We show the posterior probabilities for the Oscillatory Tracker Model (OTM) in Figure~\ref{posterior}. As we will see later, when computing the Bayesian evidence for all the $\alpha$-attractor potentials \ref{pqnA} with respect to the $o\Lambda$CDM, the OTM turns out to be the favoured model.
%As an example of the posterior results, we show the posterior for the Oscillatory Tracker Model in Figure~\ref{posterior}, for which
We have run chains of $10^5$ steps in the space of the cosmological parameters~\eqref{cosmoparam} (a four-dimensional space parameter in the case of the $\alpha$-attractors). We find that, while the standard cosmological parameters for the $o\Lambda$CDM model are well constrained to the current known values, the $\alpha$-attractor parameters present flat posteriors, which means that specific values of both, $\alpha\, ,$ and $c$ are equally likely over the explored range of the parameter space. The results obtained for all the $\alpha$-attractor potentials~\eqref{gen_potential} show this same feature,  indicating that effectively, there is a broad range parameter values in $(\alpha,c)$ for which such potentials are consistent with the set of data presented in Section~\ref{SMOS}. In fact, at least at the level of the posteriors distributions, we have not noticed significant differences among all the $\alpha$-attractor models featured. Conversely, all of them are in good agreement with the set of observations considered. This is precisely the main motivation for computing the evidence for each $\alpha$-attractor model, and the Bayes factor with respect to the $o\Lambda$CDM model.

Table~\ref{evi_bf} shows the evidence and the Bayes factor for each of the $\alpha$-attractor models we have studied.

\begin{table}[h!]\centering
\begin{tabular}{lcrcc}
\hline
\hline
\addlinespace[0.1cm]
\multirow{2}[3]{*}{\textbf{Potential}} & \multicolumn{2}{c}{\small{\textbf{Local Obs}}} & \multicolumn{2}{c}{\small{\textbf{Local Obs+PC2018}}} \\
\cmidrule(lr){2-3} \cmidrule(lr){4-5}
 & $\log \mathcal{E}$ & 2$\log B$ &$\log \mathcal{E}$ & 2$\log B$ \\
 \addlinespace[0.1cm]
\hline
\hline
\addlinespace[0.2cm]
$o\Lambda$CDM& $-528.30$ & $0$  & $-542.29$ & $0$  \\
\addlinespace[0.2cm]
L-Model& $-528.47$ & $-0.33$  & $-543.37$ & $-2.17$  \\
\addlinespace[0.2cm]
Oscillatory Tracker Model& $-527.89$ & $0.82$  & $-541.89$ & $0.80$ \\
\addlinespace[0.2cm]
Recliner Model& $-527.92$ & $0.78$  & $-542.35$ & $-0.13$  \\
\addlinespace[0.2cm]
Starobinsky Model& $-528.37$ & $-0.13$  & $-542.66$ & $-0.74$  \\
 \hline
\end{tabular}
\caption{Evidences and Bayes factor for the different $\alpha$-attractor potentials. The first two columns are the results obtained when considering only local observations (SNe Ia+OHD+BAO), whereas the last two columns show the results once adding the Planck Compressed 2018 (PC2018) data.}
\label{evi_bf}
\end{table}

\begin{figure}[h!]
\centering
\includegraphics[scale=1.]{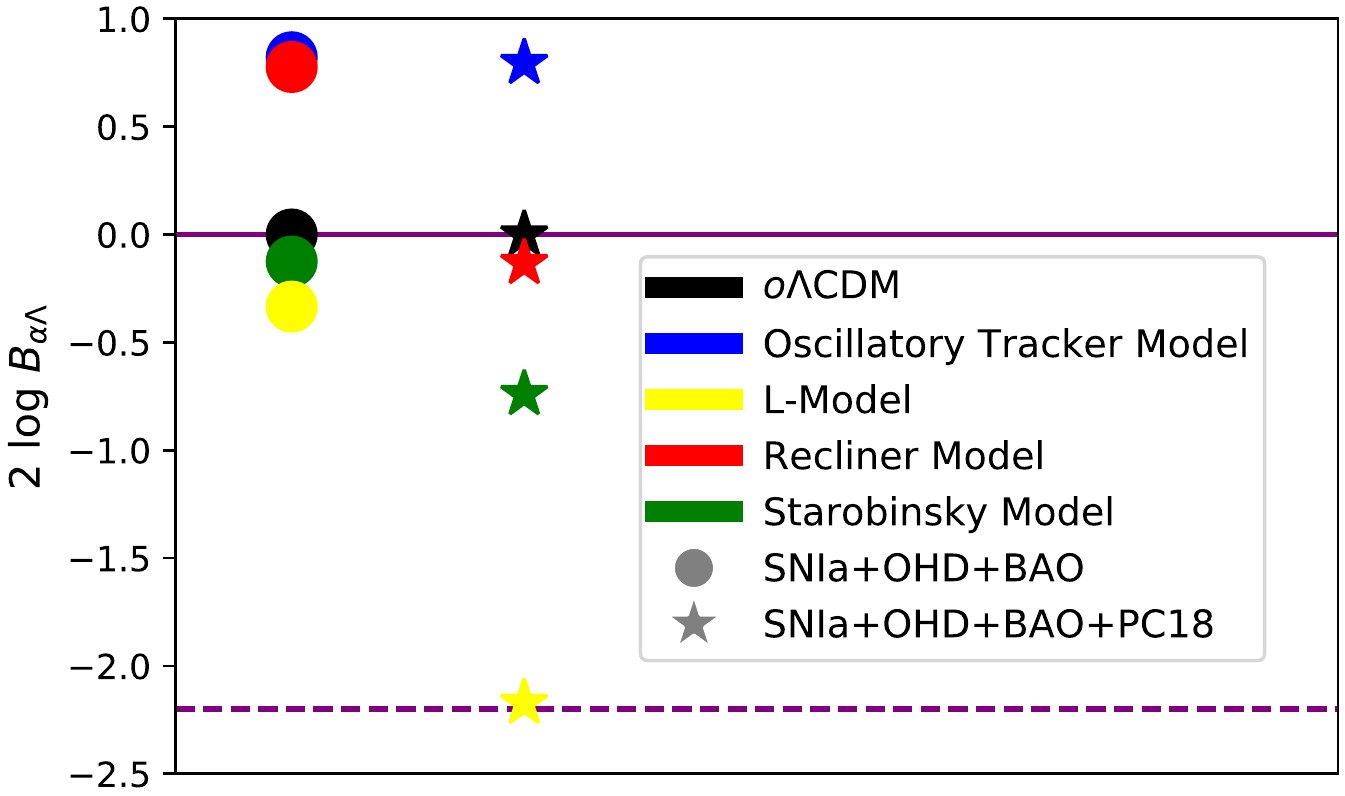}
\caption{Bayes Factor $(\log B)$ for the comparison between the $\alpha$-attractor models and $\Lambda$CDM. The first column with circles corresponds to the Bayes factor considering only local observations (SNe Ia+OHD+BAO), whereas the second column with stars indicate the Bayes factor once the Planck Compressed 2018 (PC2018) data were included. The horizontal purple solid (dashed) line labels the zero (-2.2) in the Jeffrey's scale according to Table~\ref{Jeffreyscale}. }
\label{bf_plot}
\end{figure}

When considering only local observations, the evidence indicates that the Oscillatory Tracker Model and the Recliner model are preferred over the standard cosmological model $o\Lambda$CDM. The Starobinsky potential and the L-Model are less favored, although their differences with respect $o\Lambda$CDM are small, which is expected from models mimicking the cosmological constant behavior at late times. Then, when including the Planck Compressed 2018 (PC2018) data these results are modified, and the Oscillatory Tracker Model is the only one preferred over $o\Lambda$CDM. Figure~\ref{bf_plot} displays the Bayes factors of Table~\ref{evi_bf} distinguishing pre and post Planck values in two columns. The horizontal solid line marks the zero of the Bayes factor, as $2\log B_{\alpha \Lambda}$, while the dotted line shows the threshold from models weakly disfavoured according to Jeffrey's scale.

The inclusion of the PC2018 data imposes an additional, stronger constraint on the models, and this is precisely why the Bayes factor is reduced for all the $\alpha$-attractor potentials. Following the notation used in Figure~\ref{bf_plot}, let us define the variation of the Bayes factor for each model as $\Delta \mathcal{B}_{{\rm{model}}} = \mathcal{B}_{{\rm{model}}}^{\star} - \mathcal{B}_{{\rm{model}}}^{\bullet} $, where $\mathcal{B} = 2\log B\, .$ This will allow us to measure the change on the Bayes factor from considering a set of observations or another, as a way to quantify the information added by the Planck dataset. The L-model (LM) is the one which presents the larger variation, with $\Delta \mathcal{B}_{{\rm{LM}}} = -1.84\, ,$ followed by the Recliner model (RM) with $\Delta \mathcal{B}_{{\rm{RM}}} =- 0.91\, ,$ and then the Starobinsky model (SM) $\Delta \mathcal{B}_{{\rm{SM}}} = -0.61\, .$ All of these models have a lower Bayes factor after the inclusion of PC2018. In fact, the Bayes factor for the RM was weakly preferred over $o\Lambda$CDM when considering only local observations, but disfavoured once PC2018 is added to the dataset of observations. In constrast to the above trend, the Oscillatory Tracker Model (OTM) presents a minor variation of $\Delta \mathcal{B}_{{\rm{OTM}}} = -0.02\, $. Even when considering the stronger constraint imposed by PC2018, this model is still preferred over $o\Lambda$CDM. Nonetheless, according to the Jeffrey's scale (see Table~\ref{Jeffreyscale}) such preference is not significant.

%%%%%%%%%%%%%%%%%%%%
\section{Conclusions}\label{disc}
%%%%%%%%%%%%%%%%%%%%
Bayesian evidence provides a sophisticated statistical tool for model selection in light of large amount of data. In this work we employed the \textsc{MCEvidence} code to estimate the Bayesian evidence for the $\alpha$-attractor Dark Energy models.

We have studied a variety of potentials of the $\alpha$-attractor model in light of the latest releases of cosmological data (SNe Ia, BAO, OHD and CMB observations). Our analysis represents a step beyond the parameter estimation,  already  reported for some of the realizations of this model \cite{Garcia-Garcia:2018hlc}. We have shown how  the potentials listed in Table \ref{BF} can be grouped under a single parametrisation for the $\alpha$-attractor models (see Eq.~\eqref{gen_potential}). Considering a unified form of the potential provides an advantage for model comparison since  confidence intervals for different realizations take place in a common parameter-space.

For our analysis we used two separate datasets (SNe Ia+OHD+BAO and \\SNe Ia+OHD+BAO+PC2018), finding significant improvement in the constraints on the parameter $\Omega_k$ when PC2018 were included. 

A standard statistical analysis, based on the $\chi^2$ and the Bayesian Evidence comparison through the Jeffrey's scale, shows that the Starobinsky, Recliner and L-model are moderately disfavoured by data, compared to $o\Lambda$CDM. At the same time, the Oscillatory Tracker Model is weakly favoured by the present distance observations. The relative difference in the Bayes factor for the latter with respect to the L-model indicates that the Oscillatory Tracker Model is positively favoured, advocating mainly for the consideration of this particular realization in further analyses of $\alpha$-attractor Dark Energy models. 

\section*{Acknowledgements}
FXLC thanks the Instituto de Ciencias F\'isicas at Universidad Nacional Aut\'onoma de M\'exico (ICF-UNAM) for its kind hospitality during the development of this work, and the joint support by CONACyT and DAIP-UG. A.M. is grateful to Antonio Cuesta for the generous help in the implementation of Planck compressed data in \textsc{monte python}.  The work of A.M. is supported by the DGAPA-UNAM postdoctoral grants program.  
This work is supported by the following grants: CONACyT, CB-2016-282569 and PAPIIT-UNAM, IN104119, \textit{Estudios en gravitaci\'on y cosmolog\'ia}.

\appendix

\section{$p$-value estimation for the $\alpha$-attractor models}

The use of the $p$-value as a tool to test null hypotheses is common in the literature  \cite{10.2307/2289131,berger1987,doi:10.1890/13-0590.1,hobson2014bayesian,Trotta:2017wnx}. The $p$-value is the probability that the value of some test statistic (T-statistic) be as large as or larger than the observed value assuming the null hypothesis is true\footnote{Such quantity is usually implemented in frequentist analysis, and it can be linked to the differences of the Akaike's Information Criterion ($\Delta$AIC), as well as to sigma-levels of confidence intervals~\cite{doi:10.1890/13-0590.1}.}. Moreover, a relationship between the $p$-value and an upper bound for the Bayes factor $\bar{B}$ has been previously put forward~\cite{Gordon:2007xm,Trotta:2008qt,doi:10.1198/000313001300339950}
\begin{equation}
    B \leq \bar{B} = -e \, p\log(p)\, ,\quad {\rm{for}}\quad p\leq \frac{1}{e}\, ,
    \label{pv}
\end{equation}
where $e$ is the exponential of one and $B=H_1/H_0$ is the Bayes factor, with $H_0$ ($H_1$) the null (alternative) hypothesis. Calibration of the $p$-value allows to state a scale of strength of the null hypothesis being true, as shown in Table~\ref{pvcal}, where  the lower the $p$-value, the stronger is the preference for the null hypothesis.
\begin{table}[h!]
\centering
\begin{tabular}{ccc}
\hline
\hline
$p$    & $\bar{B}$ & Strength (in favour of the null hypothesis) \\
\hline
\hline
$1/e \, \simeq $ 0.37    & 1 &       \\
%0.2 & 0.875 &        \\
%0.1 & 0.626 &      \\
0.05  & 0.407    & Weak at best     \\
0.01 & 0.125 &  \\
0.006 & 0.083 & Moderate at best \\
%0.005 & 0.072 &  \\
0.003 & 0.047 & \\
%0.001 & 0.019 &  \\
0.0003 & 0.007 & Strong at best \\
\hline
\end{tabular}
\caption{Relationship between the $p$-value and the upper bound of the Bayes factor $\bar{B}$ given by Eq.~\eqref{pv}. (See ~\cite{Trotta:2008qt,doi:10.1198/000313001300339950}).}
\label{pvcal}
\end{table}

It is our purpose in this Appendix to derive the $p$-value for the $\alpha$-attractor models. Specifically, we  consider the models given in Table~\ref{pqnA}. Based on the fact that the Bayes factor analysis we performed favours the Oscillatory Tracker Model, we pick it as the null hypothesis. By doing so, we will be able to assess the strength of our previous result (see Figure~\ref{bf_plot}) through the $p$-value. 

We present the $p$-values for the featured models in Table~\ref{bfpv}. As expected, all the $p$-values obtained have a strength lying within the ``weak at best" range, which is consistent with the weak evidence we found in the Bayesian analysis. Again, the Oscillatory Tracker Model remains a favoured model over the $o\Lambda$CDM as well as over the other $\alpha$-attractor potentials when including the Planck Compressed 2018 data. In particular, the L-Model is the least favoured of the $\alpha$-attractor models with a $p$-value given by $p<0.05$ (when Planck data is considered).
\begin{table}[h!]\centering
\begin{tabular}{lcccc}
\hline
\hline
\addlinespace[0.1cm]
\multirow{2}[3]{*}{} & \multicolumn{2}{c}{\small{\textbf{Local Obs}}} & \multicolumn{2}{c}{\small{\textbf{Local Obs+PC2018}}} \\
\cmidrule(lr){2-3} \cmidrule(lr){4-5}
 & $B$ & $p$-value &$B$ & $p$-value \\
 \addlinespace[0.1cm]
\hline
\hline
\addlinespace[0.2cm]
$o\Lambda$CDM& $0.664 $ & $0.111$  & $0.670$ & $0.113$  \\
\addlinespace[0.2cm]
L-Model& $0.560$ & $0.083$  & $0.228$ & $0.022$  \\
\addlinespace[0.2cm]
Recliner Model& $0.970$ & $0.282$  & $0.631$ & $0.102$  \\
\addlinespace[0.2cm]
Starobinsky Model& $0.619$ & $0.098$  & $0.463$ & $0.061$  \\
 \hline
\end{tabular}
\caption{Bayes factor considering as null hypothesis the Oscillatory Tracker Model, and their corresponding $p$-values for the different $\alpha$-attractor potentials and for the $o\Lambda$CDM. As in Table~\ref{evi_bf}, the first two columns show results considering only local observations (SNe Ia+OHD+BAO), whereas the last two columns show results which include the Planck Compressed 2018 (PC2018) data.}
\label{bfpv}
\end{table}

In this sense, the results we obtained in the context of $\alpha$-attractor models when testing the strength of the Oscillatory Tracker Model as the favoured model through the $p$-value, is a consistent analysis to that of the Bayesian evidence presented in Section~\ref{results}.

\bibliographystyle{plunsrt}
\bibliography{references}
\end{document}